%% file: ceccopieri.tex
\documentclass[final,3p,times,twocolumn]{elsarticle}
 \biboptions{comma,sort&compress}
\usepackage{here}
 \usepackage{graphicx}
  \usepackage{epsfig}

\def\nin{\noindent}
\def\beq{\begin{equation}}
\def\eeq{\end{equation}}
\def\bea{\begin{eqnarray}}
\def\eea{\end{eqnarray}}

\journal{Nuc. Phys. (Proc. Suppl.)}

\begin{document}

\begin{frontmatter}



\title{Associated production of one particle and a Drell-Yan pair}

\author[label1]{Federico Alberto Ceccopieri}
\address[label1]{Vrije Universiteit Brussels, 
Interfaculty Institute for High Energy (IIHE)\\
Pleinlaan 2, 1040 Brussels, Belgium.}  
\ead{federico.alberto.ceccopieri@cern.ch}

\begin{abstract}

\noindent
We briefly discuss the collinear factorization formula 
for the associated production of one particle and a Drell-Yan pair 
in hadronic collisions. 
We outline possible applications of the results to three different research 
areas.  
\end{abstract}


\end{frontmatter}


\section{Introduction}
\label{intro}
\nin
The description of particle production in hadronic collisions 
is interesting and challenging in many aspects. Perturbation 
theory can be applied whenever a sufficiently hard scale characterizes
the scattering process. 
The comparison of early LHC charged particle spectra with 
next-to-leading order perturbative QCD predictions~\cite{stratmann}
shows that the theory offers a rather good description of data at 
sufficiently high hadronic transverse momentum, of the order 
of a few GeV. For inelastic scattering processes at even lower transverse
momentum, the theoretical description in terms of perturbative QCD breaks 
down since partonic matrix elements diverge as the transverse momenta 
of final state parton vanish.
The situation is further complicated by the occurence of additional 
parton-parton interactions~\cite{DPS} accompanying the main hard process.
This effects is possibly ascribed to the composite and extended 
structure of colliding hadrons. 
The nature of the latter phenomena, which at present require 
some kind of modelization, could be more precisely investigated 
if an improved theoretical description of the main hard process were provided. 
A well suited process for these kind of studies 
is represented by the associated production of one particle and a
Drell-Yan pair, $ H_1 + H_2 \rightarrow H + \gamma^* + X$.
While the high mass lepton pair constitutes the perturbative
trigger which guarantees the applicability of perturbative QCD, 
the detected hadron $H$ could then be used, without any phase space restriction,
as a local probe to investigate particle production mechanisms. 
Within this particular process we will review how to
improve the theoretical description to make it capable to handle
the aforementioned collinear singularities associated to final state parton production 
at vanishing transverse momentum.  
We further note that this process is the single particle counterpart 
of electroweak-boson plus jets associated production~\cite{zjets}, 
presently calculated at nex-to-leading
order accuracy with up to three jets in the final state~\cite{w3jet}.
One virtue of jet requirement is that it indeed avoids 
the introduction of fragmentation functions to model the final state, 
which are instead one of the basic ingredients entering our formalism.
At variance with our case, however, jet reconstructions at very 
low transverse momentum starts to be challenging~\cite{cacciari}
and it makes difficult the study of this interesting portion of the produced particle spectrum.  

\section{Collinear facrotization formula}
\begin{figure}[h]
\centerline{\includegraphics[width=4.5cm]{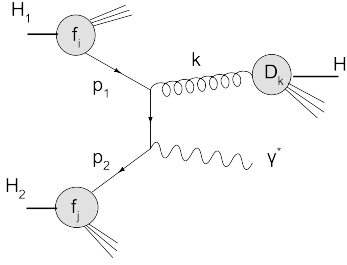}}
\caption{Example of diagram contributing to hadron production in the central
fragmentation region to order $\mathcal{O}(\alpha_s)$ in eq.(\ref{NLOcentral}).}
\label{fig1}
\end{figure}

\begin{figure}[t]
\centerline{\includegraphics[width=3.5cm]{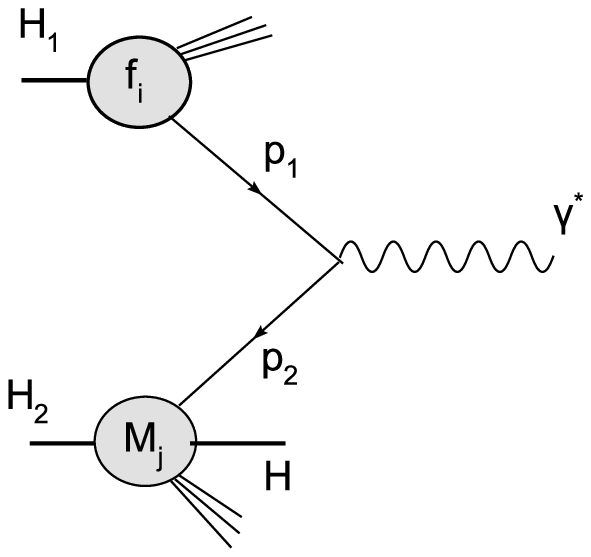}
\includegraphics[width=3.5cm]{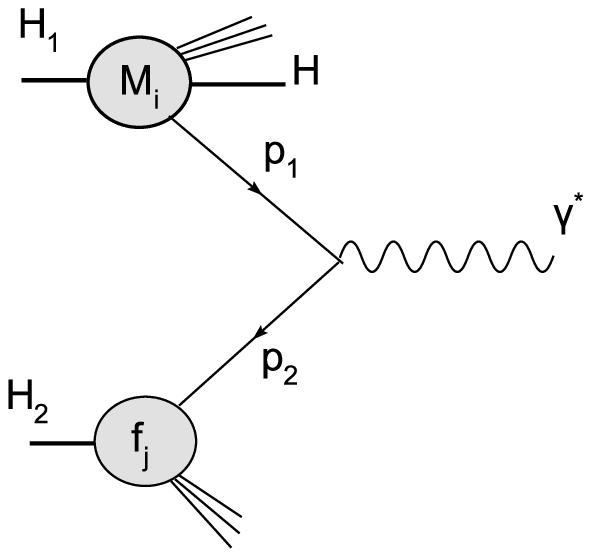}}
\caption{Parton model formula, eq.(\ref{LO}), for the associated production of a particle and a Drell-Yan pair.}
\label{fig2}
\end{figure}

\begin{figure}[t]
\centerline{\includegraphics[width=3.5cm]{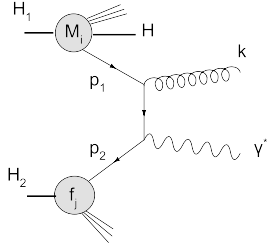}}
\caption{Example of diagram contributing to $\mathcal{O}(\alpha_s)$ corrections in the 
target fragmentation region, eq.(\ref{NLOtarget}).}
\label{fig3}
\end{figure}

\nin
The associated production of a particle and a Drell-Yan pair 
in term of partonic degrees of freedom starts 
at $\mathcal{O}(\alpha_s)$. One of the contributing diagrams 
is depicted in Fig.~(\ref{fig1}). 
Assuming that the hadronic cross-sections 
admit a factorization in term of long distance non-perturbative 
distributions and short distance perturbative calculable matrix elements, 
predictions based on perturbative QCD are obtained 
convoluting the relevant partonic sub-process cross-sections, $d\hat{\sigma}_{ij}^{k}$, 
with parton distribution functions, $f_i$ and $f_j$, and fragmentation functions,
$D^{H/k}$. The hadronic cross-sections, 
at center of mass energy squared $S=(H_1+H_2)^2$, 
can be symbolically written as~\cite{DYmio}
\beq
\label{NLOcentral}
\frac{d\sigma^{H,C,(1)}}{dQ^2 dz} \propto \frac{\alpha_s}{2\pi} \; \sum_{i,j,k} f_i^{[1]} \otimes f_{j}^{[2]} \otimes D^{H/k} 
\otimes d\hat{\sigma}_{k}^{ij},
\eeq
\nin
where the convolution are over the momentum fractions  
of the incoming and outgoing partons and $z$ is the energy of the observed hadron $H$
scaled down by the beam energy, $\sqrt{S}/2$, in the hadronic centre of mass
system. The invariant mass of the virtual photon is indicated by $Q^2$. 
The partonic indeces $i$, $j$ and $k$ in the sum run on the available partonic sub-process. 
The superscripts label the incoming hadrons and the presence of crossed term is understood.
Within this production mechanism, the observed hadron $H$ is 
generated by the fragmentation  of the final state parton $k$, and for this reason 
we address it as \textit{central}. The amplitudes squared~\cite{DYNLO}, however,
are singular when the transverse momentum of the final state parton vanishes. 
In such configurations, the parent parton $k$ of the
observed hadron $H$ is collinear the incoming parton $p_1$ or $p_2$.
Perturbation theory looses its predictivity as these phase space 
region are approached.
The same pattern of collinear singularities are found also 
in an analogue calculation in Deep Inelastic Scattering~\cite{Graudenz}. 
In both processes such singularities can not be treated with the usual 
renormalization procedure which amounts to reabsorb  collinear divergences 
into a redefinition of bare parton and fragmentation functions. 
Such singularities will appear in every fixed order calculation
in the same kinematical limits spoiling the convergence of the perturbative 
series. One possible solution is to regularize the partonic cross-sections
with a lower transverse momentum cut-off~\cite{MPI}. It would be, however, highly desiderable 
to develop a technique to resum such logarithmic contributions to all orders in perturbation
theory. 
As shown in Ref.~\cite{DYmio}, a proper treatment of such singularities 
requires the introduction of new non-perturbative distributions,
fracture functions~\cite{Trentadue_Veneziano}, $M^{H/H_1}_i(x,z)$, which parametrize hadron production 
at vanishing transverse momentum. They express the conditional probability 
to find a parton $i$ entering the hard scattering while an hadron $H$ 
is produced with fractional momentum $z$ in the target fragmentation region 
of the incoming hadron. 
By using the same generalized collinear subtraction
procedure developed for fracture functions in Ref.~\cite{Graudenz},
it is then possible to show that additional singularities discussed above 
can be properly factorized into the redefinition of bare fracture functions~\cite{DYmio}.
The fact that the collinear subtraction used in Drell-Yan process is the same as in DIS 
supports both the universality of collinear singularities among these processes
with a different number of hadrons in the initial state 
and the factorization formula in eq.~(\ref{LO}). 
Fracture functions inhomogeneous evolution equations~\cite{Trentadue_Veneziano}, 
which can be readily derived starting from this generalized collinear subtraction,  
can then be used to resum to all orders such type of collinear singularities.
The use of fracture functions opens the possibility to have 
particle production already to $\mathcal{O}(\alpha_s^0)$, 
since the hadron $H$ can be non-pertubatively produced 
by a fracture function $M$ itself.
The lowest order parton model formula can be symbolically written as
\beq
\label{LO}
\frac{d\sigma^{H,T,(0)}}{dQ^2 dz} \propto \sum_{i,j}
\big[M_i^{[1]} \otimes f_j^{[2]} + M_i^{[2]} \otimes f_j^{[1]}\big]
\eeq 
where the superscripts indicate from which incoming hadron, $H_1$ or $H_2$,
the outgoing hadron $H$ is produced through a fracture functions.
This production mechanism is sketched in Fig.~(\ref{fig2}). 
So far we have only considered $\mathcal{O}(\alpha_s)$ corrections 
in the central fragmentation region, eq.~(\ref{NLOcentral}), 
to the parton model formula, eq.~(\ref{LO})\,.
In order to complete the calculation to $\mathcal{O}(\alpha_s)$ 
we should also consider higher order corrections in process initiated by 
a fracture functions. In this case, in fact, the hadron $H$ is already produced by these 
distributions and therefore final state parton in real emission diagrams should be 
integrated over and results added to virtual corrections.
One of the contributing diagrams is depicted in Fig.~(\ref{fig3}).
The general structure of these terms is  
\beq
\label{NLOtarget}
\hspace{-0.5cm} \frac{d\sigma^{H,T,(1)}}{dQ^2 dz} \propto \frac{\alpha_s}{2\pi} \sum_{i,j}
\big[ M_i^{[1]} \otimes f_j^{[2]} + M_i^{[2]} \otimes f_j^{[1]}\big] \otimes d\hat{\sigma}^{ij}
\eeq
We refer to this corrections term as to the \textit{target} fragmentation contribution.
The calculation is, a part from different kinematics, completely  
analogue to the inclusive Drell-Yan case. While soft singularities cancel 
among real and virtual diagrams, the collinear singularities appearing in this
term are renormalized by the homogenous renormalization term in bare fracture functions and 
by bare parton distributions functions.
Adding all the various contributions, the resulting $p_t$-integrated 
cross-sections is then infrared finite~\cite{DYmio} and can be 
sintetically written as   
\bea
\label{NLOtot}
\frac{d\sigma^H}{dQ^2 dz} \propto \sum_{i,j}
\big[M_i^{[1]} \otimes f_j^{[2]} + (1 \leftrightarrow 2)\big]
\big(1+\frac{\alpha_s}{2\pi} C^{ij}\big)\nonumber\\
+\frac{\alpha_s}{2\pi} \; \sum_{i,j,k} f_i^{[1]} \otimes f_{j}^{[2]} \otimes D^{H/k} 
\otimes K_{k}^{ij}.
\eea  
We refer to the previous equation as to the collinear factorization formula 
for the process under study.
The next-to-leading coefficients $C^{ij}$ and $K_{k}^{ij}$ have been calculated~\cite{DYmio2}, 
making the whole calculation ready for numerical implementation.  
In the following section we will try to outline some applications 
of the proposed formalism. 

\section{Applications}
\nin
\textit{a)} The first possible benchmark process for the calculation 
could be strangeness production associated with a Drell-Yan pair, 
$p+p\rightarrow V+\gamma^*+X$, where $V$ generically indicates 
a $\Lambda^0$ or $\bar{\Lambda}^0$ hyperon.
The first process, according to so called leading particle effect, 
should be sensitive to the spectator system fragmentation into
$\Lambda^0$ hyperon at very low transverse momentum and therefore to be modelled with the help 
of fracture functions in $d\sigma^{H,T}$. 
In the anti-hyperon production case no such effect should be present  
so that  $\bar{\Lambda}^0$ are expected to be mainly produced by 
the fragmentation of final state parton as described by $d\sigma^{H,C}$.
The measurament of such process for different $Q^2$
could allow to test the fracture functions scale dependence embodied in their 
peculiar evolutions equations and the validity of the proposed
factorization formula in all its components. 
With this respect, a comparison of strange particle production in hadronic collisions and DIS 
would be extremely interesting since longitudinal momentum spectra of 
$\Lambda$ and $\bar{\Lambda}^0$ hyperon
has been measured quite accurately in charged current neutrino DIS~\cite{NOMAD},
both in the current as well in the target fragmentation region.

\textit{b)} The formalism may found application in the study of  
single diffractive hard process, $p+p\rightarrow p +\gamma^*+X$, 
where the outgoing proton has almost the incoming proton energy and 
extremely low transverse momentum with respect to the collision axis. 
This process has been intensively analyzed in the DIS at HERA, 
revealing its leading twist nature. From scaling violations of the diffractive 
structure functions~\cite{H106LRG} and dijet production in the final state~\cite{H107dijet,ZEUS09final} 
quite precise diffractive parton distributions functions have been extracted from data, which 
parametrize the parton content of the color singlet exchanged in the $t$-channel.
The predictions for single diffrative hard processes measured at Tevatron, 
based on diffractive parton distributions measured at HERA and assuming factorization,  
has indeed revealed that these processes are significantly suppressed in hadronic collisions, 
see the very recent analysis reported in Ref.~\cite{klasen}.
Recalling that these distributions are fracture functions in the $z\rightarrow 1$ limit, 
the present formalism can then applied to next-to-leading order accuracy, the main 
contribution to the cross-sections coming from the target term, $d\sigma^{H,T}$.
Such term can be eventually recast in triple differential form in
$x_{IP} \simeq 1-z$,  virtual photon rapidity, $y$, and invariant mass $Q^2$
and evaluated at next-to-leading order by using the appropriate coefficient 
functions~\cite{sutton}.
In this way factorization tests could be porformed at fixed $x_{IP}$ 
to avoid any Regge factorization assumption on DPDF while 
the $y$ dependence,  giving direct access to the fractional parton momentum 
in the diffractive exchange, $\beta$, 
allows to test factorization in a kinematic region which avoids 
DPDF extrapolation. Finally, the $Q^2$ dependence of the cross-sections could be used 
to investigate how factorization breaking effects eventually evolve with the hardness of the
probe and to which extent the factorized formula $M \otimes f$ actually works.        

\textit{c)} The calculation has been performed to make predictions for 
cross-sections integrated over partonic transverse momentum.
To this end, divergent contributions, related to parton emissions 
at vaninshing transverse momentum, are factorized into fracture functions. 
In the case, however, that cross-sections are measured down to
a minimum but still perturbative hadronic transverse momentum, the latter 
constitutes a natural infrared regulator for the partonic matrix elements. 
The central production term,  $d\sigma^{H,C,(1)}$, 
can be used to estimate hadron production as the fragmentation process, 
parametrized by fragmentation functions, were happening in the QCD vacuum. 
The hadronic cross-sections can be recast in a triple differential 
form in $Q^2$, produced hadron transverse momentum $p_t$ and pseudo-rapidity $\eta$
to predict charged particle spectra or multiplicity. 
A particular interesting observable which can also be reconstructed is the 
differential cross-sections differential in $\cos \phi$, 
where $\phi$ is the angle formed by the virtual photon and the detected hadron 
in the center of mass system. This observable has been shown to be sensitive 
to the contamination of the so-called underlying event~\cite{pedestal} to jet observable 
and has been used also to investigate underlying event properties 
in Drell-Yan process~\cite{UEDY_CDF}.
Although predictions made on the present formalism take into account the radiation 
accompanying one single hard scattering per proton-proton interactions, it 
can be nevertheless used as a reference
cross-sections to gauge the impact of new phenomena. 
The $d\sigma^{H,C}$ term, altough formally $\mathcal{O}(\alpha_s)$, 
is a tree level predictions and possible large higher order corrections may be 
expected especially in the forward region at large transverse momentum, 
as already found in DIS analysis~\cite{NNLO_vs_H1}.
We finally note that the recent analysis of underlying event structure in Drell-Yan events 
is performed in a wide electroweak boson mass window~\cite{UEDY_CDF}. We emphasize 
instead that more insight on the underlying event dynamics could be accessed 
if cross-sections could be measured, as in the single diffractive case discussed above,
at different virtual photon virtualities $Q^2$  and transverse momentum $q_t^2$.

\section{Conclusions}
\nin
We have briefly reviewed a perturbative approach to single particle production 
associated with a Drell-Yan pair in hadronic collisions. 
On the theoretical side we have shown that the introduction of 
new non-perturbative distributions, fracture functions, allows a consistent factorization 
of new class of collinear singularities stemming for configurations 
in which the parent parton of the observed hadron is collinear to the incoming parton.
The scale dependence induced by this generalized factorization is driven by Altarelli-Parisi
inhomogeneous evolution equations for fracture functions which allow
the resummation to all orders of this new class of collinear logarithms.  
The factorization procedure does coincide with the one used in 
DIS confirming, as expected, the universal structure of collinear singularities among 
different hadron initiated processes and supporting the collinear factorization formula 
proposed in eq.~(\ref{NLOtot}).   
On the phenomenological side we have briefly discussed a few applications 
in which different aspects of the formalism could be tested
and compared towards results already obtained in the DIS target fragmentation region.
A good theoretical control on the perturbative component indeed allows the investigation 
on new phenomena appearing in hadronic collisions, for example the rapidity gap probability 
suppression in hard diffractive processes with respect to diffractive DIS and 
the investigation, although indirect, of multiple parton-parton interactions. 

\section*{Acknowledgements}
\nin
The author would like to thank S. Narison for kind invitation to this 
stimulating conference. 











\input{bib_ceccopieri}

\end{document}